\documentclass[seceq]{ptptex}
\usepackage{graphicx, wrapft, braket}
\input{colordvi.tex}

\title{CMB Bispectrum from Primordial Scalar, Vector and Tensor Non-Gaussianities}

\author{Maresuke \textsc{Shiraishi}, Daisuke \textsc{Nitta}, Shuichiro \textsc{Yokoyama}, Kiyotomo \textsc{Ichiki} and Keitaro \textsc{Takahashi}}
\inst{Department of Physics and Astrophysics, Nagoya University, Nagoya 464-8602, Japan}

\abst{
We present an all-sky formalism for the Cosmic Microwave Background (CMB) bispectrum induced by the
primordial non-Gaussianities not only in scalar but also in vector and tensor fluctuations.
We find that the bispectrum can be
formed in an explicitly rationally invariant way  
by taking into account 
the angular and polarization dependences of the vector
 and tensor modes. To demonstrate this and present how to use
 our formalism, we consider a specific example of the correlation between two scalars and a
 graviton as the source of non-Gaussianity. As a result, we show that
 the CMB reduced bispectrum of the intensity anisotropies is evaluated
 as a function of the multipole and the coupling constant between two scalars and a graviton denoted by $g_{tss}$; $|b_{\ell \ell \ell}| \sim \ell^{-4} \times 8 \times 10^{-18} |g_{tss}|$. By estimating the signal-to-noise ratio, we find
 that the constraint as $|g_{tss}| < 6$ will be expected from the PLANCK
 experiment.}

\begin{document}
\maketitle

\section{Introduction}

Bispectrum (three-point correlation functions) of the Cosmic Microwave
Background (CMB) temperature anisotropies has been attracting attention
over the years as a powerful observational tool to investigate the
primordial non-Gaussianity \cite{Komatsu:2001rj,Bartolo:2004if}.  As is
well known, primordial curvature perturbations whose statistics deviate
from the pure Gaussian ones can produce the nonzero bispectrum of the
CMB temperature anisotropies.  The size of the primordial
non-Gaussianity has often been parametrized by a so-called nonlinearity
parameter $f_{\rm NL}$.  Depending on the shape of the bispectrum, this
nonlinearity parameter can be sorted into three types of $f_{\rm NL}$,
called local, equilateral and orthogonal types.  Current
observational limits on these $f_{\rm NL}$s are given by $ -10 < f_{\rm
NL}^{\rm local} < 74 $ for the local type, $ -214 < f_{\rm NL}^{\rm
equil} < 266 $ for the equilateral type and $ - 410< f_{\rm NL}^{\rm
orthog} < 6 $ for the orthogonal type (95\,\% C.L.) \cite{Komatsu:2010fb}. 
These observational constraints are still consistent with 
the Gaussian primordial curvature perturbations that are expected to be generated from the standard single slow-roll inflation
model.  However, one can also consider that there may be some cue for
hunting the non-Gaussianity because the central values of some $f_{\rm
NL}$s have  deviated from zero.  Hence, if future experiments
would confirm that the statistics of the primordial curvature
perturbations deviate from the Gaussian ones, then the standard
single slow-roll inflation model can be excluded as a dominant mechanism
of generating primordial fluctuations.  Thus, the primordial
non-Gaussianity can be considered as a new probe of the mechanism of
generating primordial curvature perturbations.

As discussed above, previous studies have focused largely on non-Gaussianities in scalar-mode perturbations. However, we can also consider the sources of non-Gaussianities in vector and tensor perturbations, for example, the nonlinear couplings between gravitons and scalars during inflation \cite{Maldacena:2002vr}, nonlinearities of the Sachs-Wolfe effect~\cite{Mollerach:1997up,Bartolo:2003bz,Bartolo:2005kv,Boubekeur:2009uk}, cosmic strings \cite{Takahashi:2008ui,Hindmarsh:2009qk}, and primordial magnetic fields \cite{Brown:2005kr,Seshadri:2009sy,Caprini:2009vk,Cai:2010uw,Trivedi:2010gi,Shiraishi:2010yk,Kahniashvili:2010us}. Hence, in order to constrain the size of non-Gaussianities and understand the nature of these sources, one should include the contribution of vector and tensor non-Gaussianities to the CMB bispectrum. 

In our previous work~\cite{Shiraishi:2010sm},
we presented the bispectrum formulae of the CMB
temperature and polarization anisotropies sourced from non-Gaussianity
not only in scalar but also in vector and tensor fluctuations.  
It is expected that the nonlinearities will also induce
the tensor and vector mode
perturbations, and the modes may generate more characteristic features
in the CMB angular spectra than in the scalar one. 
In Ref.~\citen{Shiraishi:2010sm}~,  
we found that the bispectrum formulae for vector and
 tensor modes in all sky analysis formally take complicated forms
 compared with the scalar mode case owing to the dependence of the photon
 transfer functions on the azimuthal angle between the wave vector
 of photon fluctuation $\mib{k}$ and the unit vector specifying the line of sight
 direction $\hat{\mib{n}}$. 
However, by using flat sky approximation, we have simplified
 the equations of bispectra of the CMB anisotropies to solve the above
 difficulty because no azimuthal dependence arises in this limit.

This paper is an extension of our previous work.  We present a general
formalism of the CMB bispectrum induced from the primordial vector and
tensor fluctuations in the all-sky analysis. We newly consider the angular
dependences in the polarization vector and tensor bases, which have been
neglected in our previous work. To demonstrate how to
calculate the CMB bispectrum by making use of our formula, we show
a calculation of the CMB bispectrum induced from the primordial
non-Gaussianity generated through the interaction between two scalars
and a graviton (tensor) during inflation, which has been originally
discussed by Maldacena~\cite{Maldacena:2002vr}.

This paper is organized as follows.
In the next section, we present a formulation of the CMB
bispectrum in the all-sky approach,
which is an extension of our previous work~\cite{Shiraishi:2010sm}.
In \S\ref{sec:maldacena},
we show the calculation of the CMB bispectrum induced from the
primordial non-Gaussianity generated through the two scalars and a
graviton (tensor) correlator during inflation. We define a coupling parameter
characterizing the strength of such interaction as $g_{tss}$ and
evaluate an observational limit on $g_{tss}$ by calculating the signal-to-noise ratio.
In the final section, we give a summary and conclusion of this paper. 

\section{Formulation of the CMB bispectra for scalar, vector and tensor
 modes}
\label{sec:formula}

In this section, we derive general formulae of the CMB bispectrum of
temperature and polarization fluctuations induced by the primordial
non-Gaussianity in scalar, vector or tensor-mode perturbations in the
all-sky analysis.

At first, we introduce an expression of CMB
fluctuation. In the all-sky analysis, CMB fluctuations of intensity or
polarization field are expanded with the spin-0 or spin-2 spherical
harmonics, respectively \cite{Shiraishi:2010sm, Zaldarriaga:1996xe, Lewis:2004ef}. Then, the coefficients of CMB fluctuations, called
$a_{\ell m}$, are described as
\begin{eqnarray}
a_{X, \ell m}^{(Z)} &=& 4\pi (-i)^\ell \int \frac{d^3 \mib{k}}{(2 \pi)^3} 
\sum_\lambda [{\rm sgn}(\lambda)]^{\lambda+x} {}_{-\lambda}Y^*_{\ell
m}(\hat{\mib{k}}) \xi^{(\lambda)}(\mib{k}) \mathcal{T}^{(Z)}_{X,\ell}(k)~, \label{eq:alm_ang}
\end{eqnarray}
where the index $Z$ denotes the mode of perturbations: $Z = S$
(scalar), $= V$ (vector) or $=T$ (tensor) and its helicity is expressed
by $\lambda$; $\lambda = 0$ for $(Z = S)$, $= \pm 1$ for $(Z = V)$ or $=
\pm 2$ for $(Z = T)$, $X$ discriminates between intensity and two
polarization (electric and magnetic) modes, respectively, as $X = I,E,B$
and $x$ is determined by it: $x = 0$ for $X = I,E$ or $=1$ for $X = B$,
$\xi^{(\lambda)}$ is the initial perturbation decomposed on each helicity state and $\mathcal{T}_{X, \ell}^{(Z)}$ is the time-integrated transfer function in each
sector (calculated in, for example, Refs.~\citen{Shiraishi:2010sm, Landriau:2002fx, Hu:1997hp}).\footnote{Here, we set $0^0 = 1$.}

Next, we expand $\xi^{(\lambda)}$ with spin-$(-\lambda)$ spherical harmonics as
\begin{eqnarray}
\xi^{(\lambda)}(\mib{k}) \equiv \sum_{\ell m} \xi^{(\lambda)}_{\ell m} (k)
{}_{-\lambda} Y_{\ell m}(\hat{\mib{k}})~, \label{eq:xi_def}
\end{eqnarray}
 and eliminate the angular dependence in Eq.~(\ref{eq:alm_ang}) by performing
$\hat{\mib{k}}$-integral: 
\begin{eqnarray}
a^{(Z)}_{X, \ell m} &=& 4 \pi (-i)^\ell \int_0^\infty \frac{k^2 dk}{(2 \pi)^3}
\sum_\lambda [{\rm sgn}(\lambda)]^{\lambda+x} \xi_{\ell m}^{(\lambda)}(k) 
\mathcal{T}^{(Z)}_{X, \ell}(k)~. \label{eq:alm}
\end{eqnarray}
Here, we use the orthogonality relation of spin-$\lambda$ spherical
harmonics as \cite{Newman/Penrose:1966, Goldberg/etal:1967}
\begin{eqnarray}
\int d^2 \hat{\mib{n}} {}_{\lambda}Y^*_{\ell' m'}(\hat{\mib{n}})
{}_{\lambda}Y_{\ell m}(\hat{\mib{n}}) = \delta_{\ell, \ell'}\delta_{m, m'}~.
\end{eqnarray}

The initial bispectrum in vector or tensor-mode perturbations will be expressed as
\begin{eqnarray}
\Braket{\prod_{i=1}^3 \xi^{(\lambda_i)}(\mib{k_i}) }
\equiv (2 \pi)^3 
F^{\lambda_1 \lambda_2 \lambda_3}(\mib{k_1}, \mib{k_2}, \mib{k_3})
\delta \left( \sum_{i=1}^3 \mib{k_i} \right)~.
\label{eq:initial_bis_ang}
\end{eqnarray} 
This definition, which includes the angular dependence on $\mib{k}$ in the
polarization vector or tensor, is more general than Eq.~(3) of
Ref.~\citen{Shiraishi:2010sm}. 
We will see in the
discussion in \S\ref{sec:maldacena} that the
initial bispectrum from inflation 
indeed takes the above form for the tensor case. 
On the other hand, to calculate the CMB bispectrum using
Eq.~(\ref{eq:alm}), the bispectrum of $\xi_{\ell m}$ is needed. If the
primordial bispectrum satisfies the rotational invariance, we can set it as
\begin{eqnarray}
\Braket{\prod_{i=1}^3 \xi^{(\lambda_i)}_{\ell_i m_i}(k_i) } 
\equiv (2 \pi)^3 \mathcal{F}_{\ell_1 \ell_2
\ell_3}^{\lambda_1 \lambda_2 \lambda_3} (k_1, k_2, k_3)
\left(
  \begin{array}{ccc}
  \ell_1 & \ell_2 & \ell_3 \\
  m_1 & m_2 & m_3 
  \end{array}
 \right)~. \label{eq:initial_bis}
\end{eqnarray}
Then, using Eq.~(\ref{eq:xi_def}), the conversion equation between
$F^{\lambda_1 \lambda_2 \lambda_3}$ and $\mathcal{F}^{\lambda_1
\lambda_2 \lambda_3}_{\ell_1 \ell_2, \ell_3}$ is derived as 
\begin{eqnarray}
\mathcal{F}_{\ell_1 \ell_2 \ell_3}^{\lambda_1 \lambda_2 \lambda_3} (k_1, k_2, k_3) 
&=& \sum_{m_1 m_2 m_3} 
\left(
  \begin{array}{ccc}
  \ell_1 & \ell_2 & \ell_3 \\
  m_1 & m_2 & m_3 
  \end{array}
 \right) 
\left[ \prod_{i=1}^3 \int d^2 \hat{\mib{k_i}} {}_{-\lambda_i} Y^*_{\ell_i
 m_i}(\hat{\mib{k_i}}) \right] \nonumber \\
&&\times 
F^{\lambda_1 \lambda_2
\lambda_3}(\mib{k_1}, \mib{k_2}, \mib{k_3})
\delta\left( \sum_{i=1}^3 \mib{k_i} \right)~. \label{eq:F_conv}
\end{eqnarray}
From Eqs.~(\ref{eq:alm}), (\ref{eq:initial_bis}) and the orthogonality
of Wigner-$3j$ symbols as Eq.~(\ref{eq:Wig_3j_lllmmm}), 
the CMB angle-averaged bispectrum, which is defined as \cite{Komatsu:2001rj, Bartolo:2004if}
\begin{eqnarray}
B_{X_1 X_2 X_3 , \ell_1,\ell_2,\ell_3}^{(Z_1 Z_2 Z_3)}
&\equiv& \sum_{m_1 m_2 m_3} 
\left(
  \begin{array}{ccc}
  \ell_1 & \ell_2 & \ell_3 \\
  m_1 & m_2 & m_3 
  \end{array}
 \right) 
\Braket{\prod_{i=1}^3 a^{(Z_i)}_{X_i, \ell_i m_i}}~,
\end{eqnarray}
can be written as 
\begin{eqnarray}
B_{X_1 X_2 X_3 , \ell_1,\ell_2,\ell_3}^{(Z_1 Z_2 Z_3)} 
&=& \left[ \prod_{n=1}^3 4\pi (-i)^{\ell_n} \int_0^\infty \frac{k_n^2 dk_n}{(2 \pi)^3} 
\mathcal{T}^{(Z_n)}_{X_n, \ell_n}(k_n) 
\sum_{\lambda_n} [{\rm sgn}(\lambda_n)]^{\lambda_n + x_n}
\right] 
\nonumber \\
&& \times
(2 \pi)^3
\mathcal{F}_{\ell_1 \ell_2 \ell_3}^{\lambda_1 \lambda_2 \lambda_3} (k_1, k_2, k_3)~.
\label{eq:cmb_bis}
\end{eqnarray}
Thus, when one computes the CMB bispectrum, only the alternative initial
bispectrum $\mathcal{F}^{\lambda_1 \lambda_2 \lambda_3}_{\ell_1 \ell_2
\ell_3}$ is necessary in each case.
\section{CMB bispectrum induced by the primordial non-Gaussianity in the two
scalars and a graviton correlator}\label{sec:maldacena}

In this section, we demonstrate how to calculate $\mathcal{F}^{\lambda_1
\lambda_2 \lambda_3}_{\ell_1 \ell_2 \ell_3}$ and the CMB bispectrum by
considering the contribution of two scalars and a graviton correlator
\cite{Maldacena:2002vr}. Furthermore, we evaluate an observational limit
on the primordial non-Gaussianity of the graviton sector by calculating the
signal-to-noise ratio.

\subsection{Two scalars and a graviton interaction during inflation}

We consider a general single-field inflation model with Einstein-Hilbelt action
\cite{Garriga:1999vw}~:
\begin{eqnarray}
S=\int d^4x \sqrt{-g}\left[\frac{M_{\rm pl}^2}{2}R
+p(\phi,X)\right] ~,
\end{eqnarray}
where $g$ is the determinant of the metric, $R$ is the Ricci scalar, $M_{\rm
pl}^2\equiv 1/(8\pi G)$, $\phi$ is a scalar field, and $X \equiv - g^{\mu \nu}
\partial_{\mu}\phi\partial_\nu \phi/2$. Using the background equations, the
slow-roll parameter and the sound speed for perturbations are given by
\begin{eqnarray}
\epsilon \equiv -\frac{\dot{H}}{H^2}=\frac{X p_{,X}}{H^2M_{\rm pl}^2}~, \ \
c_s^2 \equiv \frac{p_{,X}}{2 X p_{,XX} + p_{,X}} ~,
\end{eqnarray}
where $H$ is the Hubble parameter, the dot means a derivative with respect to the
physical time $t$ and $p_{,X}$ denotes partial derivative of $p$ with respect to
$X$.
 We write a metric by ADM formalism
\begin{eqnarray}
ds^2 = -N^2dt^2+a^2e^{\gamma_{ab}}
(dx^a+N^adt)(dx^b+N^bdt) ~,
\end{eqnarray}
where $N$ and $N^a$ are respectively the lapse function and shift vector,
$\gamma_{ab}$ is a transverse and traceless tensor as
$\gamma_{aa}=\partial_a\gamma_{ab}=0$, and $e^{\gamma_{ab}} \equiv
\delta_{ab}+\gamma_{ab}+\gamma_{ac}\gamma_{cb}/2+\cdots$. On the 
flat hypersurface, the gauge-invariant curvature perturbation
$\zeta$ is related to the first-order fluctuation of the scalar field
$\varphi$ as $\zeta = - H\varphi/\dot{\phi}$.  
Following the conversion equations (\ref{eq:scal_decompose}) and (\ref{eq:tens_decompose}), we decompose $\zeta$ and $\gamma_{ab}$ into the helicity states as 
\begin{eqnarray}
\xi^{(0)}(\mib{k}) = {\zeta}(\mib{k}) ~ , \ \
\xi^{(\pm 2)}(\mib{k}) &=& \frac{1}{2} e_{ab}^{(\mp 2)}(\hat{\mib{k}}) \gamma_{ab}(\mib{k}) ~.
\end{eqnarray}
Here, $e^{(\pm 2)}_{ab}$ is a transverse and traceless polarization
tensor explained in Appendix \ref{appen:polarization}.  
The interaction
parts of this action have been derived by Maldacena
\cite{Maldacena:2002vr} up to the third-order terms.  
In particular, we
will focus on an interaction between two scalars and a graviton.
This is because the correlation  between a small wave number of the tensor
mode and large wave numbers of the scalar modes will remain despite the
tensor mode decays after the mode reenters the cosmic horizon.
We find a leading term of the two scalars
and a graviton interaction in the action coming from the matter part of
the Lagrangian through $X$ as
\begin{eqnarray}
X|_{\rm 3rd-order}\supset a^{-2}\frac{p_{,X}}{2}\gamma_{ab}\partial_a \varphi 
\partial_b\varphi~,
\end{eqnarray}
therefore, the interaction part is given by
\begin{eqnarray}
S_{\rm int}\supset\int d^4x\, a g_{tss} \gamma_{ab}\partial_a \zeta \partial_b\zeta~.
\end{eqnarray}
Here, we introduce a coupling constant $g_{tss}$. 
From the definition of $\zeta, \gamma_{ab}$ and the slow-roll parameter, $g_{tss} = \epsilon$.  For a general
 consideration, let us deal with $g_{tss}$ as a free parameter.  In
 this sense, constraining on this parameter may offer a probe of the
 nature of inflation and gravity in the early universe.  The primordial
 bispectrum is then computed using in-in formalism in the next
 subsection.

\subsection{Calculation of the initial bispectrum}

\begin{figure}[t]
  \centering \includegraphics[width=10cm,height=7cm,clip]{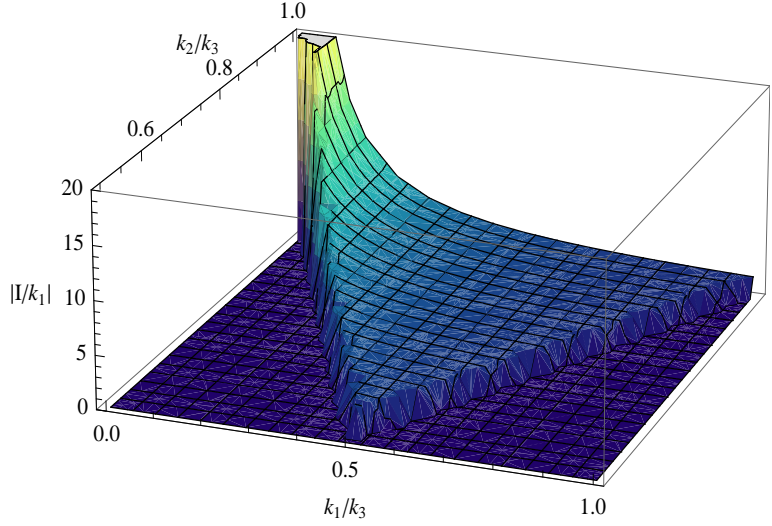}
  \caption{(color online) Shape of 
$I / k_1$. For the symmetric property and the triangle
 condition, we limit the plot range as $k_1 \leq k_2 \leq k_3$ and $|k_1 - k_2| \leq k_3 \leq k_1 + k_2$.}
  \label{fig:SST_Iok1}
\end{figure}

In the same manner as discussed in Ref.~\citen{Maldacena:2002vr}, we
calculate the primordial bispectrum generated from two scalars and a
graviton in the lowest order of the slow-roll parameter:
\begin{eqnarray}
\Braket{\xi^{(\pm 2)}(\mib{k_1}) \xi^{(0)}(\mib{k_2}) \xi^{(0)}(\mib{k_3})}
 &=&
(2 \pi)^3 \delta(\mib{k_1} + \mib{k_2} + \mib{k_3})
\frac{4 g_{tss} I(k_1,k_2,k_3) k_2 k_3}{\prod_i
(2k^3_i)}\frac{H_*^4}{2c^2_{s*}\epsilon^2_* M_{\rm pl}^4} \nonumber \\
&&\times e_{ab}^{(\mp 2)}(\hat{\mib{k_1}}) \hat{k_2}_a \hat{k_3}_b ~, \label{eq:mal_bis} \\
I(k_1, k_2, k_3) &\equiv&  -k_t + \frac{k_1 k_2 + k_2 k_3 + k_3
 k_1}{k_t} + \frac{k_1 k_2 k_3}{k_t^2}~,
\end{eqnarray}
where $k_t \equiv k_1 + k_2 + k_3$, and $*$ means that it is evaluated
at the time of horizon crossing, i.e., $a_*H_* = k$.  Here, we keep the
angular and polarization dependences, $e_{ab}^{(\mp 2)}(\hat{\mib{k_1}})
\hat{k_2}_a \hat{k_3}_b$, which have sometimes been omitted in the
literature for simplicity \cite{Andrew/etal:2002,Shiraishi:2010sm,Caprini:2009vk}. 
We show, however, that expanding this
term with spin-weighted spherical harmonics enables us to formulate the
rotational-invariant bispectrum in an explicit way.
The statistically isotropic power spectra of $\xi^{(0)}$ and
$\xi^{(\pm 2)}$ are respectively given by
\begin{eqnarray}
\Braket{\xi^{(0)}(\mib{k}) \xi^{(0) *}(\mib{k'})}
&\equiv& (2\pi)^3 P_S(k) \delta(\mib{k} - \mib{k'})~, \nonumber \\ 
\frac{k^3 P_S(k)}{2 \pi^2} &=& \frac{H_*^2}{8 \pi^2 c_{s*}\epsilon_* M_{\rm pl}^2} \equiv A_S~, \nonumber \\
\Braket{\xi^{(\lambda)}(\mib{k}) \xi^{(\lambda') *}(\mib{k'})}
&\equiv& (2\pi)^3 \frac{P_T(k)}{2} \delta(\mib{k} - \mib{k'})
\delta_{\lambda, \lambda'} \ ({\rm for} \  \lambda = \pm 2)
~, \nonumber \\
\frac{k^3 P_T(k)}{2 \pi^2} &=& \frac{H_*^2}{ \pi^2 M_{\rm pl}^2} 
= 8 c_{s *}
 \epsilon_* A_S \equiv \frac{r}{2} A_S~, \label{eq:def_power}
\end{eqnarray}
where $r$ is the tensor-to-scalar ratio 
and $A_S$ is the amplitude of primordial curvature perturbations. 
Note that the power spectra satisfy the scale invariance because we consider them in the lowest order of the slow-roll parameter.
Using these equations, we parametrize the initial bispectrum in this case from
Eqs.~(\ref{eq:mal_bis}) and (\ref{eq:initial_bis_ang}) as
\begin{eqnarray}
F^{\pm2 0 0}(\mib{k_1}, \mib{k_2}, \mib{k_3}) 
&=& f^{(TSS)}(k_1, k_2, k_3) e_{ab}^{(\mp 2)}(\hat{\mib{k_1}}) \hat{{k_2}}_a
\hat{k_3}_b~, \label{eq:mal_F} \\
f^{(TSS)}(k_1, k_2, k_3) &\equiv& \frac{16 \pi^4 A_S^2 g_{tss}}{k_1^2 k_2^2 k_3^2} \frac{I(k_1, k_2, k_3)}{k_t} \frac{k_t}{k_1} ~. \label{mal_ftss}
\end{eqnarray}
Note that $f^{(TSS)}$ seems not to depend on the tensor-to-scalar
ratio. 
In Fig.~\ref{fig:SST_Iok1}~, we show the shape of $I/k_1$. From this, we confirm that the initial bispectrum $f^{(TSS)}$ (\ref{mal_ftss}) dominates in the squeezed limit as $k_1 \ll k_2 \simeq k_3$ like the local-type bispectrum of scalar modes.

In the squeezed limit, the ratio of $f^{(TSS)}$ to the
scalar-scalar-scalar counterpart $f^{(SSS)} = \frac{6}{5}f_{\rm NL}
P_S(k_1) P_S(k_2)$, which has been considered
frequently, reads
\begin{equation}
\frac{f^{(TSS)}}{f^{(SSS)}}
= \frac{10 g_{tss}}{3 f_{\rm NL}} \frac{I}{k_t} \frac{k_t k_2}{k_3^2}
\rightarrow \frac{20 g_{tss}}{3 f_{\rm NL}} \frac{I}{k_t}~.
\end{equation}
In the standard slow-roll inflation model, this ratio becomes ${\cal
O}(1)$ and does not depend on
the tensor-to-scalar ratio because $g_{tss}$ and $f_{\rm NL}$ are
proportional to the slow-roll parameter $\epsilon$, and 
$I/k_t$ has a nearly flat shape. The average of amplitude is evaluated as
$I/k_t \approx -0.6537$. Therefore, it manifests the comparable
importance of the higher order correlations of tensor modes to the scalar ones in the standard inflation scenario.

\subsection{Formulation of the CMB bispectrum}

For this case, by substituting Eq.~(\ref{eq:mal_F}) into
Eq.~(\ref{eq:F_conv}), the initial bispectrum
is given by 
\begin{eqnarray}
\mathcal{F}_{\ell_1 \ell_2 \ell_3}^{\pm2 0 0} (k_1, k_2, k_3) 
&=& \sum_{m_1 m_2 m_3} 
\left(
  \begin{array}{ccc}
  \ell_1 & \ell_2 & \ell_3 \\
  m_1 & m_2 & m_3 
  \end{array}
 \right)
\left(\prod_{i=1}^3 \int d^2 \hat{\mib{k_i}} \right) \nonumber \\
&& \times
{}_{\mp 2}Y^*_{\ell_1 m_1}(\hat{\mib{k_1}}) Y^*_{\ell_2 m_2}(\hat{\mib{k_2}}) Y^*_{\ell_3
m_3}(\hat{\mib{k_3}})  \nonumber \\
&& \times 
f^{(TSS)}({k_1},
 {k_2}, {k_3}) e_{ab}^{(\mp 2)}(\hat{\mib{k_1}}) \hat{k_2}_a \hat{k_3}_b
\delta\left( \sum_{i=1}^3 \mib{k_i} \right) ~.
\label{eq:mal_calF}
\end{eqnarray}
We derive this simpler form as the following procedure.

At first, we express all parts containing the angular dependence with
the spin spherical harmonics:
\begin{eqnarray}
e^{(\mp 2)}_{ab}(\hat{\mib{k_1}})\hat{k_2}_a \hat{k_3}_b 
&=& \frac{4 (8\pi)^{3/2}}{3} \sum_{M m_a m_b}
{}_{\pm 2} Y^*_{2 M}(\hat{\mib{k_1}})
Y^*_{1 m_a}(\hat{\mib{k_2}}) Y^*_{1 m_b}(\hat{\mib{k_3}}) 
\left(
  \begin{array}{ccc}
  2 &  1 & 1 \\
   M & m_a & m_b
  \end{array}
 \right)~, \nonumber \\ \\
\delta\left( \sum_{i=1}^3 \mib{k_i} \right) 
&=& 8 \int_0^\infty y^2 dy 
\left[ \prod_{i=1}^3 \sum_{L_i M_i} 
 (-1)^{L_i/2} j_{L_i}(k_i y) 
Y_{L_i M_i}^*(\hat{\mib{k_i}}) \right] \nonumber \\
&&\times 
I_{L_1 L_2 L_3}^{0~0~0}
\left(
  \begin{array}{ccc}
  L_1 & L_2 & L_3 \\
  M_1 & M_2 & M_3 
  \end{array}
 \right)~, \label{eq:delta}
\end{eqnarray}
where we used the relations listed in  Appendices \ref{appen:wigner} and \ref{appen:polarization} and 
\begin{eqnarray}
I^{s_1 s_2 s_3}_{l_1 l_2 l_3}
\equiv \sqrt{\frac{(2 l_1 + 1)(2 l_2 + 1)(2 l_3 + 1)}{4 \pi}}
\left(
  \begin{array}{ccc}
  l_1 & l_2 & l_3 \\
  s_1 & s_2 & s_3
  \end{array}
 \right)~.
\end{eqnarray} 
Secondly, using Eq.~(\ref{eq:gaunt}), we replace all the integrals of spin spherical
harmonics with the Wigner symbols:
\begin{eqnarray}
\int d^2 \hat{\mib{k_1}}~ {}_{\mp 2}Y_{\ell_1 m_1}^*(\hat{\mib{k_1}}) 
Y_{L_1 M_1}^*(\hat{\mib{k_1}})  {}_{\pm 2}Y_{2 M}^*(\hat{\mib{k_1}})
&=& I^{\pm 2 0 \mp 2}_{\ell_1 L_1 2}
\left(
  \begin{array}{ccc}
  \ell_1 &  L_1 & 2 \\
   m_1 & M_1 & M
  \end{array}
 \right)~, \\
\int d^2 \hat{\mib{k_2}}~ Y_{\ell_2 m_2}^*(\hat{\mib{k_2}}) 
Y_{L_2 M_2}^*(\hat{\mib{k_2}}) Y_{1 m_a}^*(\hat{\mib{k_2}}) 
&=& I^{0~ 0~ 0}_{\ell_2 L_2 1}
\left(
  \begin{array}{ccc}
  \ell_2 &  L_2 & 1 \\
   m_2 & M_2 & m_a
  \end{array}
 \right)~, \\
\int d^2 \hat{\mib{k_3}}~ Y_{\ell_3 m_3}^*(\hat{\mib{k_3}}) 
Y_{L_3 M_3}^*(\hat{\mib{k_3}}) Y_{1 m_b}^*(\hat{\mib{k_3}})
&=& I^{0~ 0~ 0}_{\ell_3 L_3 1}
\left(
  \begin{array}{ccc}
  \ell_3 &  L_3 & 1 \\
   m_3 & M_3 & m_b
  \end{array}
 \right)~.
\end{eqnarray}  
Thirdly, using the summation formula of five Wigner-$3j$ symbols as Eq.~(\ref{eq:sum_9j_3j}), we sum up the Wigner-$3j$ symbols
with respect to azimuthal quantum numbers in the above equations and express
with the Wigner-$9j$ symbol as
\begin{eqnarray}
&& \sum_{\substack{M_1 M_2 M_3 \\ M m_a m_b}}
\left(
  \begin{array}{ccc}
  L_1 &  L_2 & L_3 \\
   M_1 & M_2 & M_3
  \end{array}
 \right)
\left(
  \begin{array}{ccc}
  2 &  1 & 1 \\
   M & m_a & m_b
  \end{array}
 \right) \nonumber \\
&&\qquad \times
\left(
  \begin{array}{ccc}
  \ell_1 &  L_1 & 2 \\
   m_1 & M_1 & M
  \end{array}
 \right)
\left(
  \begin{array}{ccc}
  \ell_2 &  L_2 & 1 \\
   m_2 & M_2 & m_a
  \end{array}
 \right)
\left(
  \begin{array}{ccc}
  \ell_3 &  L_3 & 1 \\
   m_3 & M_3 & m_b
  \end{array}
 \right) \nonumber \\
&& \qquad\qquad\qquad = 
\left(
  \begin{array}{ccc}
  \ell_1 & \ell_2 & \ell_3 \\
   m_1 & m_2 & m_3
  \end{array}
 \right)
\left\{
  \begin{array}{ccc}
  \ell_1 & \ell_2 & \ell_3 \\
   L_1 & L_2 & L_3 \\
   2 & 1 & 1 \\
  \end{array}
 \right\}~.
\end{eqnarray}
After these treatments, performing the summation over
$m_1,m_2$ and $m_3$ like Eq.~(\ref{eq:Wig_3j_lllmmm}), 
we can obtain the final form as
\begin{eqnarray}
\mathcal{F}_{\ell_1 \ell_2 \ell_3}^{\pm 2 0 0} (k_1, k_2, k_3) 
&=& \frac{(8\pi)^{3/2}}{6}
f^{(TSS)}({k_1}, {k_2}, {k_3}) 
\nonumber \\
&& \times
\sum _{L_1 L_2 L_3} I_{L_1 L_2 L_3}^{0~0~0} 
I^{\pm 2 0 \mp 2}_{\ell_1 L_1 2}
I^{0~ 0~ 0}_{\ell_2 L_2 1} I^{0~ 0~ 0}_{\ell_3 L_3 1} 
\left\{
  \begin{array}{ccc}
  \ell_1 & \ell_2 & \ell_3 \\
   L_1 & L_2 & L_3 \\
   2 & 1 & 1 \\
  \end{array}
 \right\} \nonumber \\
&&\times
\int_0^\infty y^2 dy 
\left[ \prod_{i=1}^{3}
(-1)^{L_i/2}  j_{L_i} (k_i y) \right]
~. \label{eq:mal_calF_final}
\end{eqnarray}
Note that the absence of the summation over $m_1, m_2$ and $m_3$ in this equation means that the tensor-scalar-scalar bispectrum maintains the
rotational invariance. As described above, this consequence is derived from
the angular dependence in the polarization tensor. Also in vector modes,
if their power spectra obey the statistical isotropy like
Eq.~(\ref{eq:def_power}), one can obtain the rotational invariant bispectrum by
considering the angular dependence in the polarization vector 
as Eq.~(\ref{eq:pol_vec}).

Then, substituting the expression (\ref{eq:mal_calF_final}) into Eq.~(\ref{eq:cmb_bis}), we can calculate the CMB bispectrum induced from the nonlinear coupling between two scalars and a graviton. 
The CMB angle-averaged bispectrum is derived as 
\begin{eqnarray}
&&B^{(TSS)}_{X_1 X_2 X_3, \ell_1 \ell_2 \ell_3} \nonumber \\
&&\quad= \frac{(8 \pi)^{3/2}}{3} \sum_{L_1 L_2 L_3} 
(-1)^{\frac{L_1 + L_2 + L_3}{2}} I^{0~0~0}_{L_1 L_2 L_3} 
I^{2 0 -2}_{\ell_1 L_1 2} I^{0~0~0}_{\ell_2 L_2 1} I^{0~ 0~ 0}_{\ell_3 L_3 1} 
\left\{
  \begin{array}{ccc}
  \ell_1 & \ell_2 & \ell_3 \\
  L_1 & L_2 & L_3 \\
  2 & 1 & 1
  \end{array}
 \right\} \nonumber \\
&&\qquad \times \int_0^\infty y^2 dy  
\left[ \prod_{n=1}^3 \frac{2}{\pi} (-i)^{\ell_n} \int_0^\infty k_n^2 d
 k_n {\cal T}_{X_n, \ell_n}^{(Z_n)} j_{L_n}(k_n y)  \right]
f^{(TSS)}(k_1, k_2, k_3)~, \nonumber \\ 
\label{eq:mal_cmb_bis}   
\end{eqnarray} 
where we use the summation over $\lambda_1 = \pm 2$ as
\begin{eqnarray}
\sum_{\lambda_1 = \pm 2} [{\rm sgn}(\lambda_1)]^{\lambda_1 + x_1} 
I^{\lambda_1 0 -\lambda_1}_{\ell_1 L_1 2} = 
\begin{cases}
2 I^{2 0 -2}_{\ell_1 L_1 2}&
({\rm for \ } x_1 + L_1 + \ell_1 = {\rm even} )~, \\
0 & ({\rm for \ } x_1 + L_1 + \ell_1 = {\rm odd} ) ~.
\end{cases}
\end{eqnarray}
Considering the selection rules of the Wigner symbols explained in Appendix
\ref{appen:wigner}, we see that the bispectrum (\ref{eq:mal_cmb_bis}) has nonzero value under the conditions:
\begin{eqnarray}
&& L_1 =
\begin{cases}
|\ell_1 \pm 2|, \ell_1 & ({\rm for \ } X_1 = I,E ) \\
|\ell_1 \pm 1| & ({\rm for \ } X_1 = B )
\end{cases}~, \ \ 
 L_2 = |\ell_2 \pm 1|~, \ \
 L_3 = |\ell_3 \pm 1|~, \nonumber \\
&& |L_1 - L_2| \leq L_3 \leq L_1 + L_2~, \ \ 
\sum_{i=1}^3 L_i = {\rm even}~, \nonumber\\
&& |\ell_1 - \ell_2| \leq \ell_3 \leq \ell_1 + \ell_2~, \ \ 
\sum_{i=1}^3 \ell_i =
\begin{cases}
{\rm even} & ({\rm for \ } X_1 = I,E ) ~, \\
{\rm odd} & ({\rm for \ } X_1 = B ) ~.
\end{cases} 
\end{eqnarray} 

\begin{figure}[t]
  \centering \includegraphics[width=10cm,height=7cm,clip]{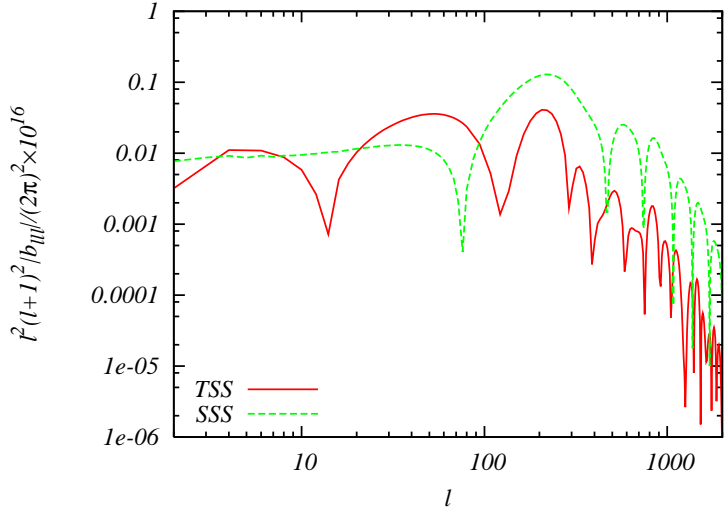}
  \caption{(color online) Absolute values of the CMB reduced bispectra of 
 temperature fluctuation for $\ell_1 = \ell_2 = \ell_3$. 
The lines correspond to the spectra 
  generated from tensor-scalar-scalar correlation given by
 Eq.~(\ref{eq:mal_cmb_red_bis}) with $g_{tss} = 5$ (red solid line) and
 the primordial non-Gaussianity in the scalar curvature perturbations
 with $f^{\rm local}_{\rm NL} = 5$ (green dashed line). 
 The other cosmological parameters are fixed to the mean values limited
 from WMAP-7yr data reported in Ref.~\citen{Komatsu:2010fb}.}
  \label{fig:SST_III_samel}
\end{figure}

\begin{figure}[ht]
  \centering \includegraphics[width=10cm,height=7cm,clip]{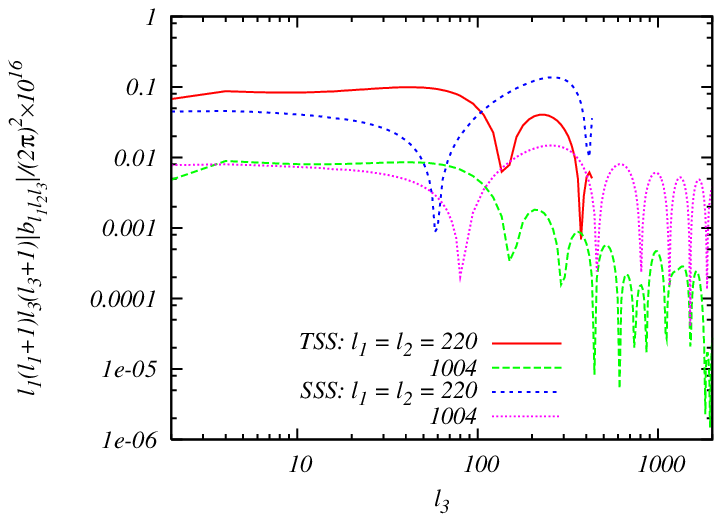}
  \caption{(color online) Absolute values of the CMB reduced bispectra of 
 temperature fluctuation generated from tensor-scalar-scalar correlation
 given by Eq.~(\ref{eq:mal_cmb_red_bis}) $(TSS)$ and the primordial non-Gaussianity in the scalar curvature perturbations $(SSS)$ as a function of $\ell_3$ with
 $\ell_1$ and $\ell_2$ fixed to some values as indicated. The
 parameters are fixed to the same values defined in
 Fig.~\ref{fig:SST_III_samel}.}
  \label{fig:SST_III_difl}
\end{figure}

In Figs.~\ref{fig:SST_III_samel} and \ref{fig:SST_III_difl}, we describe the reduced CMB bispectra of intensity mode sourced from two scalars and a graviton coupling: 
\begin{eqnarray}
&& b^{(TSS)}_{I I I, \ell_1 \ell_2 \ell_3} + b^{(STS)}_{I I I, \ell_1
 \ell_2 \ell_3} + b^{(SST)}_{I I I, \ell_1 \ell_2 \ell_3} \nonumber \\
&&\qquad = \left( I_{\ell_1 \ell_2 \ell_3}^{0~0~0}\right)^{-1}
\left(B^{(TSS)}_{I I I, \ell_1 \ell_2 \ell_3} 
+ B^{(STS)}_{I I I, \ell_1 \ell_2 \ell_3} 
+ B^{(SST)}_{I I I, \ell_1 \ell_2 \ell_3} 
\right)~,
\label{eq:mal_cmb_red_bis}
\end{eqnarray}
and primordial curvature perturbations: 
\begin{eqnarray}
b^{(SSS)}_{III, \ell_1 \ell_2 \ell_3} 
= \left(I^{0~0~0}_{\ell_1 \ell_2 \ell_3}\right)^{-1} 
B^{(SSS)}_{III, \ell_1 \ell_2 \ell_3}~.
\end{eqnarray} 
 For the numerical computation, we modify the Boltzmann Code for
Anisotropies in the Microwave Background (CAMB) \cite{Lewis:2004ef,
Lewis:1999bs}. In the calculation of the Wigner-$3j$ and $9j$ symbols,
we use the Common Mathematical Library SLATEC \cite{slatec} and the
summation formula of three Wigner-$6j$ symbols (\ref{eq:sum_9j_6j}). As
the radiation transfer functions of scalar and tensor modes, namely, ${\cal T}^{(S)}_{X_i, \ell_i}$ and ${\cal T}^{(T)}_{X_i, \ell_i}$, we
use the 1st-order formulae as discussed in Refs.~\citen{Hu:1997hp} and \citen{Lewis:1999bs}.
From the behavior of each line shown in Fig.~\ref{fig:SST_III_difl} at small $\ell_3$ that the reduced CMB bispectrum is roughly proportional to $\ell^{-2}$, we can confirm that the tensor-scalar-scalar bispectrum has a nearly squeezed-type
configuration corresponding to the shape of the initial bispectrum as discussed above. From Fig.~\ref{fig:SST_III_samel}, by comparing the green dashed line with the red solid line roughly estimated as  
\begin{eqnarray}
|b^{(TSS)}_{I I I, \ell \ell \ell} + b^{(STS)}_{I I I, \ell \ell \ell}
+ b^{(SST)}_{I I I, \ell \ell \ell}| 
\sim \ell^{-4} \times 8 \times 10^{-18}
 |g_{tss}|~, \label{eq:mal_cmb_bis_raugh}
\end{eqnarray}
we find that $|g_{tss}| \sim 5$ is comparable to
$f^{\rm local}_{\rm NL} = 5$ corresponding to the upper bound expected
from the PLANCK experiment.
In the next subsection, we check the validity of the above evaluation by computation of the signal-to-noise ratio assuming the zero-noise data.

\subsection{Estimation of the signal-to-noise ratio}

Here, we compute the signal-to-noise ratio by comparing the intensity bispectrum of
Eq.~(\ref{eq:mal_cmb_bis}) with the zero-noise (ideal) data and examine the bound on
the absolute value of $g_{tss}$. 
The formulation of (the square of) the signal-to-noise ratio $(S/N)$ is
reported in Refs.~\citen{Komatsu:2001rj} and \citen{Bartolo:2004if}. In our case, it can be
expressed as 
\begin{eqnarray}
\left( \frac{S}{N} \right)^2 = \sum_{2 \leq \ell_1 \leq \ell_2 \leq
 \ell_3 \leq \ell} 
\frac{ \left(B^{(TSS)}_{I I I \ell_1 \ell_2 \ell_3} 
+ B^{(STS)}_{I I I \ell_1 \ell_2 \ell_3} 
+ B^{(SST)}_{I I I \ell_1 \ell_2 \ell_3}\right)^2 }{\sigma^2_{\ell_1 \ell_2
\ell_3}}~, \label{eq:SN}
\end{eqnarray}
where $\sigma_{\ell_1 \ell_2 \ell_3}$ denotes the variance of the bispectrum. 
Assuming the weakly non-Gaussianity, the variance 
can be estimated as \cite{Spergel:1999xn, Gangui:2000gf} 
\begin{eqnarray}
\sigma^2_{\ell_1 \ell_2 \ell_3} \approx C_{\ell_1} C_{\ell_2} C_{\ell_3}
\Delta_{\ell_1 \ell_2 \ell_3}~, 
\end{eqnarray}
where $\Delta_{\ell_1 \ell_2 \ell_3}$ takes $1, 6$ or $2$ 
for $\ell_1 \neq \ell_2 \neq \ell_3, \ell_1 = \ell_2 = \ell_3$, or the
case that two $\ell$'s are the same, respectively. $C_\ell$ denotes that the CMB
angular power spectrum included the noise spectrum, which is
neglected in our case.   

\begin{figure}[t]
  \centering \includegraphics[width=9cm,height=6cm,clip]{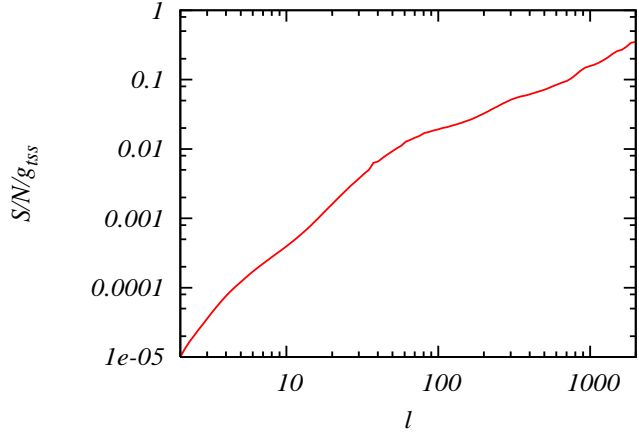}
  \caption{(color online) Signal-to-noise ratio normalized by
 $g_{tss}$ as a function
 of the maximum value between $\ell_1, \ell_2$ and $\ell_3$, namely, $\ell$. Each parameter is fixed to the same values defined in Fig. \ref{fig:SST_III_samel}.}
  \label{fig:SST_SN}
\end{figure}

In Fig. \ref{fig:SST_SN}, the numerical result of Eq.~(\ref{eq:SN}) is
presented. We find that $(S/N)$ is a monotonically increasing function
roughly proportional to $\ell$ for $\ell < 2000$. It is
compared with the order estimation of Eq.~(\ref{eq:SN}) as
Ref.~\citen{Bartolo:2004if}
\begin{eqnarray}    
\left( \frac{S}{N} \right) &\sim& \sqrt{ \frac{\ell^3}{24} } 
\times \sqrt{\frac{(2 \ell)^3}{4 \pi}}
\left| \left(
  \begin{array}{ccc}
  \ell & \ell & \ell \\
   0 & 0 & 0
  \end{array}
 \right) \right| 
\frac{\ell^3 | b^{(TSS)}_{I I I \ell \ell \ell} + b^{(STS)}_{I I I \ell \ell
\ell} + b^{(SST)}_{I I I \ell \ell \ell} |}{(\ell^2 C_\ell)^{3/2}}
\nonumber \\
&\sim& \ell \times 5.4 \times 10^{-5} |g_{tss}| ~.
\end{eqnarray}
Here, we use Eq.~(\ref{eq:mal_cmb_bis_raugh}) and the approximations as $\sum \sim \ell^3/24$, 
$\ell^3 
\left(
  \begin{array}{ccc}
  \ell & \ell & \ell \\
   0 & 0 & 0
  \end{array}
 \right)^2 \sim 0.36 \times \ell$, and $\ell^2 C_\ell \sim 6 \times
 10^{-10}$. 
 We confirm
 that this is consistent with Fig. \ref{fig:SST_SN}, which justifies
 our numerical calculation in some sense.
This figure shows that from the WMAP and PLANCK
experimental data \cite{Komatsu:2010fb, :2006uk}, 
which are roughly noise-free at $\ell \lesssim 500$ and $1000$, respectively, 
expected $(S/N) / g_{tss}$ values are $0.072$ and
 $0.16$. Hence, to obtain $(S/N) > 1$, we need
 $|g_{tss}| > 14 $ and $6$. The latter value is consistent with a naive estimate $|g_{tss}| \lesssim 5$, which was discussed at the end of the previous section.

\section{Summary and discussion }\label{sec:sum}

In this paper, we present a full-sky formalism of the CMB bispectrum
 sourced from the primordial non-Gaussianity not only in the scalar 
 but also in the vector and tensor perturbations. 
As an extension of the previous formalism discussed in
 Ref.~\citen{Shiraishi:2010sm}, the new formalism contains the
 contribution of the polarization vector and tensor in the initial
 bispectrum.  In Ref.~\citen{Shiraishi:2010sm}, we have shown that
 in the all-sky analysis, the CMB bispectrum of vector or tensor mode
 cannot be formed as a simple angle-averaged bispectrum in the same
 way as that of scalar mode.  This is because the angular integrals over
 the wave number vectors have complexities for the non-orthogonality of
 spin spherical harmonics whose spin values differ from each other if
 one neglects the angular dependence of the polarization vector or
 tensor. In this study, however, we find that this difficulty vanishes
 if we maintain the angular dependence in the initial bispectrum.
 
 To present how to use our formalism, we
 compute the CMB bispectrum induced by the nonlinear mode-coupling
 between the two scalars and a graviton \cite{Maldacena:2002vr}. The
 typical value of the reduced bispectrum in temperature fluctuations is
 calculated as a function of the coupling constant between scalars and
 gravitons $g_{tss}$: $|b^{(TSS)}_{III, \ell \ell \ell} + b^{(STS)}_{III, \ell
 \ell \ell} + b^{(SST)}_{III, \ell \ell \ell}| \sim \ell^{-4} \times 8
 \times 10^{-18} |g_{tss}|$.  Through the computation of the
 signal-to-noise ratio, we expect a constraint as $|g_{tss}| < 14 $ from
 WMAP and $|g_{tss}| < 6$ from PLANCK. Although we do not include the
 effect of the polarization modes in the estimation of $g_{tss}$ in
 this study, they will provide more beneficial information of the
 nonlinear nature of the early universe.

Our formalism will be applicable to the other sources of vector or
tensor non-Gaussianity, such as, the cosmic strings \cite{Takahashi:2008ui,
Hindmarsh:2009qk} or the primordial magnetic fields
\cite{Kahniashvili:2010us}. Actually, in the specific case of 
vector-vector-vector correlation, we have already presented the
rotationally invariant bispectrum
from primordial magnetic fields \cite{Shiraishi:2010yk}.

\section*{Acknowledgements}
This work is supported by a Grant-in-Aid for JSPS Research under Grant
 No. 22-7477 (M. S.), JSPS Grants-in-Aid for Scientific Research under Grant 
 Nos. 22340056 (S. Y.), 21740177, 22012004 (K. I.), and 21840028 (K. T.).
This work is also supported in part by the Grant-in-Aid for Scientific
Research on Priority Areas No. 467 ``Probing the Dark Energy through an
 Extremely Wide and Deep Survey with Subaru Telescope'' and by the
 Grant-in-Aid for Nagoya University Global COE Program, ``Quest for
 Fundamental Principles in the Universe: from Particles to the Solar
 System and the Cosmos,'' from the Ministry of Education, Culture,
 Sports, Science and Technology of Japan. 

\appendix
\section{Useful Properties of the Wigner Symbols}\label{appen:wigner}

Here, we briefly review the useful properties of the Wigner-$3j, 6j$ and $9j$
symbols. The following discussions are based on Refs.~\citen{Okamoto:2002ik, Gurau:2008, Jahn/Hope:1954, mathematica, Hu:2001fa}.

\subsection{Wigner-$3j$ symbol}

In quantum mechanics, considering the coupling of two angular momenta as
\begin{eqnarray}
\mib{l_3} = \mib{l_1} + \mib{l_2}~,
\end{eqnarray}
the scalar product of eigenstates between the right-handed term and the left-handed one, namely, a Clebsch-Gordan coefficient, is related to the 
Wigner-$3j$ symbol:
\begin{eqnarray}
\left(
  \begin{array}{ccc}
  l_1 & l_2 & l_3 \\
   m_1 & m_2 & - m_3
  \end{array}
 \right) 
\equiv \frac{ (-1)^{l_1 - l_2 + m_3} \Braket{l_1 m_1 l_2 m_2 |
(l_1 l_2) l_3 m_3} }{\sqrt{2 l_3 + 1}}~.
\end{eqnarray}
This symbol vanishes unless the selection rules are satisfied as follows:
\begin{eqnarray}
&& |m_1| \leq l_1~, \ \ |m_2| \leq l_2~, \ \ |m_3| \leq l_3 ~, \ \ m_1 + m_2 = m_3 ~, \nonumber \\
&& |l_1 - l_2| \leq l_3 \leq l_1 + l_2 \ {\rm (the \ triangle \ condition)}~, \ \ l_1 + l_2 + l_3 \in \mathbb{Z} ~.
\end{eqnarray}
Symmetries of the Wigner-$3j$ symbol are given by 
\begin{eqnarray}
\left(
  \begin{array}{ccc}
  l_1 & l_2 & l_3 \\
   m_1 & m_2 & m_3
  \end{array}
 \right) 
&=& (-1)^{\sum_{i=1}^3 l_i} \left(
  \begin{array}{ccc}
  l_2 & l_1 & l_3 \\
  m_2 & m_1 & m_3
  \end{array}
 \right) 
= (-1)^{\sum_{i=1}^3 l_i} \left(
  \begin{array}{ccc}
  l_1 & l_3 & l_2 \\
  m_1 & m_3 & m_2
  \end{array}
 \right) \nonumber \\ 
&&\qquad\qquad ({\rm odd \ permutation \ of \ columns}) \nonumber \\
&=& \left(
  \begin{array}{ccc}
  l_2 & l_3 & l_1 \\
  m_2 & m_3 & m_1
  \end{array}
 \right) 
= \left(
  \begin{array}{ccc}
  l_3 & l_1 & l_2 \\
  m_3 & m_1 & m_2
  \end{array}
 \right) \nonumber \\ 
&&\qquad\qquad ({\rm even \ permutation \ of \ columns}) \nonumber \\
&=& (-1)^{\sum_{i=1}^3 l_i} 
\left(
  \begin{array}{ccc}
  l_1 & l_2 & l_3 \\
  - m_1 & - m_2 & - m_3
  \end{array}
 \right) \nonumber \\
&&\qquad\qquad ({\rm sign \ inversion \ of} \ m_1, m_2, m_3) ~ .
\end{eqnarray}
The Wigner-$3j$ symbols satisfy the orthogonality as 
\begin{eqnarray}
 (2 l_3 + 1) \sum_{l_3 m_3} 
\left(
  \begin{array}{ccc}
  l_1 & l_2 & l_3 \\
  m_1 & m_2 & m_3
  \end{array}
 \right)
\left(
  \begin{array}{ccc}
  l_1 & l_2 & l_3 \\
  m_1' & m_2' & m_3
  \end{array}
 \right) 
&=& \delta_{m_1, m'_1} \delta_{m_2, m'_2}~, \nonumber \\
 (2 l_3 + 1) \sum_{m_1 m_2} 
\left(
  \begin{array}{ccc}
  l_1 & l_2 & l_3 \\
  m_1 & m_2 & m_3
  \end{array}
 \right)
\left(
  \begin{array}{ccc}
  l_1 & l_2 & l'_3 \\
  m_1 & m_2 & m'_3
  \end{array}
 \right) 
&=& \delta_{l_3, l_3'} \delta_{m_3, m'_3}~. \label{eq:Wig_3j_lllmmm}
\end{eqnarray}
For a special case that $\sum_{i=1}^3 l_i = {\rm even}$ and $m_1 = m_2 = m_3 = 0$, there is an analytical expression as 
\begin{eqnarray}
&& \left(
  \begin{array}{ccc}
  l_1 & l_2 & l_3 \\
  0 & 0 & 0
  \end{array}
 \right) \nonumber \\
&&= (-1)^{\sum_{i=1}^3 {- l_i \over 2}} 
\frac{ \left(\sum_{i=1}^3 \frac{l_i}{2} \right)! \sqrt{(-l_1 + l_2 + l_3)!}
\sqrt{(l_1 - l_2 + l_3)!} \sqrt{(l_1 + l_2 - l_3)!} }
{ \left(\frac{- l_1 + l_2 + l_3}{2}\right)! 
\left(\frac{l_1 - l_2 + l_3}{2}\right)! 
\left(\frac{l_1 + l_2 - l_3}{2}\right)! 
\sqrt{ \left( \sum_{i=1}^3 l_i + 1 \right)!} }.
\end{eqnarray}
This vanishes for $\sum_{i=1}^3 l_i = {\rm odd}$.
The Wigner-$3j$ symbol is related to the spin-weighted spherical harmonics as
\begin{eqnarray}
\prod_{i=1}^2 {}_{s_i} Y_{l_i m_i}(\hat{\mib{n}})
 &=& \sum_{l_3 m_3 s_3} {}_{s_3} Y^*_{l_3 m_3}(\hat{\mib{n}}) 
I^{-s_1 -s_2
-s_3}_{l_1~l_2~l_3} 
\left(
  \begin{array}{ccc}
  l_1 & l_2 & l_3 \\
  m_1 & m_2 & m_3
  \end{array}
 \right)~, \label{eq:product_sYlm} 
\end{eqnarray}
which leads to the ``extended'' Gaunt integral including spin dependence:
\begin{eqnarray}
\int d^2 \hat{\mib{n}} {}_{s_1} Y_{l_1 m_1}(\hat{\mib{n}}) {}_{s_2} Y_{l_2 m_2}
(\hat{\mib{n}}){}_{s_3} Y_{l_3 m_3}(\hat{\mib{n}}) 
= I^{-s_1 -s_2 -s_3}_{l_1~l_2~l_3}
\left(
  \begin{array}{ccc}
  l_1 & l_2 & l_3 \\
  m_1 & m_2 & m_3
  \end{array}
 \right)~. \label{eq:gaunt}
\end{eqnarray}
Here $I^{s_1 s_2 s_3}_{l_1 l_2 l_3}
\equiv \sqrt{\frac{(2 l_1 + 1)(2 l_2 + 1)(2 l_3 + 1)}{4 \pi}}
\left(
  \begin{array}{ccc}
  l_1 & l_2 & l_3 \\
  s_1 & s_2 & s_3
  \end{array}
 \right)$.


\subsection{Wigner-$6j$ symbol}

Considering two other ways in the coupling of three angular momenta as
\begin{eqnarray}
\mib{l_5} &=& \mib{l_1} + \mib{l_2} + \mib{l_4} \\
&=& \mib{l_3} + \mib{l_4} \label{eq:l3l4} \\
&=& \mib{l_1} + \mib{l_6}~, \label{eq:l1l6}
\end{eqnarray}
the Wigner-$6j$ symbol is defined using a Clebsch-Gordan coefficient
between each eigenstate of $\mib{l_5}$ corresponding to Eqs.~(\ref{eq:l3l4}) and (\ref{eq:l1l6}) as 
\begin{eqnarray}
\left\{
  \begin{array}{ccc}
  l_1 & l_2 & l_3 \\
  l_4 & l_5 & l_6
  \end{array}
 \right\}
 \equiv \frac{ (-1)^{l_1 + l_2 + l_4 + l_5} \Braket{(l_1 l_2) l_3 ; l_4 ;
 l_5 m_5 | l_1 ; (l_2 l_4) l_6 ; l_5 m_5} }{\sqrt{(2 l_3 + 1)(2 l_6 + 1)}}~.
\end{eqnarray}
This is expressed with the summation of three Wigner-$3j$ symbols:
\begin{eqnarray}
&& \sum_{m_4 m_5 m_6} (-1)^{\sum_{i=4}^6 l_i - m_i}
\left(
  \begin{array}{ccc}
  l_5 & l_1 & l_6 \\
  m_5 & -m_1 & -m_6 
  \end{array}
 \right) \nonumber \\
&&\qquad \times
\left(
  \begin{array}{ccc}
  l_6 & l_2 & l_4 \\
  m_6 & -m_2 & -m_4 
  \end{array}
 \right)
\left(
  \begin{array}{ccc}
  l_4 & l_3 & l_5 \\
  m_4 & -m_3 & -m_5 
  \end{array}
 \right) \nonumber \\
 &&\qquad\qquad\qquad = \left(
  \begin{array}{ccc}
  l_1 & l_2 & l_3 \\
  m_1 & m_2 & m_3 
  \end{array}
 \right) 
\left\{
  \begin{array}{ccc}
  l_1 & l_2 & l_3 \\
  l_4 & l_5 & l_6 
  \end{array}
 \right\}~; 
 \end{eqnarray}
hence, the triangle conditions are given by 
\begin{eqnarray}
&& |l_1 - l_2| \leq l_3 \leq l_1 + l_2, \ |l_4 - l_5| \leq l_3 \leq l_4 + l_5 ~,   \nonumber \\
&& |l_1 - l_5| \leq l_6 \leq l_1 + l_5, \ |l_4 - l_2| \leq l_6 \leq l_4 + l_2~. \label{eq:wig_6j_triangle}
\end{eqnarray}
The Wigner-$6j$ symbol obeys 24 symmetries such as
\begin{eqnarray}
\left\{
  \begin{array}{ccc}
  l_1 & l_2 & l_3 \\
  l_4 & l_5 & l_6 
  \end{array}
 \right\} 
&=& \left\{
  \begin{array}{ccc}
  l_2 & l_1 & l_3 \\
  l_5 & l_4 & l_6 
  \end{array}
 \right\} 
= \left\{
  \begin{array}{ccc}
  l_2 & l_3 & l_1 \\
  l_5 & l_6 & l_4 
  \end{array}
 \right\} \ ({\rm permutation \ of \ columns}) \nonumber \\
&=& \left\{
  \begin{array}{ccc}
  l_4 & l_5 & l_3 \\
  l_1 & l_2 & l_6 
  \end{array}
 \right\} 
= \left\{
  \begin{array}{ccc}
  l_1 & l_5 & l_6 \\
  l_4 & l_2 & l_3 
  \end{array}
 \right\} \nonumber \\ 
&& ({\rm exchange \ of \ two \ corresponding \ elements \ between \ rows}).
\end{eqnarray}
Geometrically, the Wigner-$6j$ symbol is expressed using the tetrahedron composed of four triangles obeying Eq.~(\ref{eq:wig_6j_triangle}). It is known that the Wigner-$6j$ symbol is suppressed by the square root of the volume of the tetrahedron at high multipoles.   

\subsection{Wigner-$9j$ symbol}

Considering two other ways in the coupling of four angular momenta as
\begin{eqnarray}
\mib{l_9} &=& \mib{l_1} + \mib{l_2} + \mib{l_4} + \mib{l_5} \\
&=& \mib{l_3} + \mib{l_6} \label{eq:l3l6} \\
&=& \mib{l_7} + \mib{l_8}~, \label{eq:l7l8}
\end{eqnarray}
where $\mib{l_3} \equiv \mib{l_1} + \mib{l_2}, {\mib l_6} \equiv \mib{l_4} + \mib{l_5}, \mib{l_7} \equiv \mib{l_1} + {\mib l_4}, \mib{l_8} \equiv \mib{l_2} + \mib{l_5}$,
the Wigner $9j$ symbol expresses a Clebsch-Gordan coefficient between
 each eigenstate of $\mib{l_9}$ corresponding to Eqs.~(\ref{eq:l3l6}) and
(\ref{eq:l7l8}) as 
\begin{eqnarray}
&& \left\{
 \begin{array}{ccc}
  l_1 & l_2 & l_3 \\
  l_4 & l_5 & l_6 \\
  l_7 & l_8 & l_9
 \end{array}
 \right\}  
\equiv \frac{\Braket{(l_1 l_2) l_3 ; (l_4 l_5) l_6 ; l_9 m_9 | (l_1
 l_4) l_7 ; (l_2 l_5) l_8 ; l_9 m_9}}{\sqrt{(2 l_3 + 1)(2 l_6 + 1)(2 l_7 + 1)(2 l_8
+ 1)}}~. 
\end{eqnarray}
This is expressed with the summation of five Wigner-$3j$ symbols:
\begin{eqnarray}
&& \sum_{\substack{m_4 m_5 m_6 \\ m_7 m_8 m_9}} 
\left(
  \begin{array}{ccc}
  l_4 & l_5 & l_6 \\
  m_4 & m_5 & m_6 
  \end{array}
 \right)
\left(
  \begin{array}{ccc}
  l_7 & l_8 & l_9 \\
  m_7 & m_8 & m_9 
  \end{array}
 \right) \nonumber \\
&&\qquad \times 
\left(
  \begin{array}{ccc}
  l_4 & l_7 & l_1 \\
  m_4 & m_7 & m_1 
  \end{array}
 \right)
\left(
  \begin{array}{ccc}
  l_5 & l_8 & l_2 \\
  m_5 & m_8 & m_2
  \end{array}
 \right)
\left(
  \begin{array}{ccc}
  l_6 & l_9 & l_3 \\
  m_6 & m_9 & m_3 
  \end{array}
 \right) \nonumber \\ 
&& \qquad\qquad\qquad = \left(
  \begin{array}{ccc}
  l_1 & l_2 & l_3 \\
  m_1 & m_2 & m_3
  \end{array}
 \right)
\left\{
  \begin{array}{ccc}
  l_1 & l_2 & l_3 \\
  l_4 & l_5 & l_6 \\
  l_7 & l_8 & l_9 
  \end{array}
 \right\}~, \label{eq:sum_9j_3j}
\end{eqnarray}
and that of three Wigner-$6j$ symbols:
\begin{eqnarray}
\left\{
 \begin{array}{ccc}
  l_1 & l_2 & l_3 \\
  l_4 & l_5 & l_6 \\
  l_7 & l_8 & l_9
 \end{array}
 \right\}
&=& \sum_x (-1)^{2x}(2 x + 1) \nonumber \\
&& \times 
\left\{
 \begin{array}{ccc}
  l_1 & l_4 & l_7 \\
  l_8 & l_9 & x 
 \end{array}
 \right\}
\left\{
 \begin{array}{ccc}
  l_2 & l_5 & l_8 \\
  l_4 & x & l_6 
 \end{array}
 \right\}
\left\{
 \begin{array}{ccc}
  l_3 & l_6 & l_9 \\
  x & l_1 & l_2 
 \end{array}
 \right\} ; \label{eq:sum_9j_6j}
\end{eqnarray}
hence, the triangle conditions are given by
\begin{eqnarray}
&& |l_1 - l_2| \leq l_3 \leq l_1 + l_2~, \ |l_4 - l_5| \leq l_6 \leq l_4 +
 l_5~, \ |l_7 - l_8| \leq l_9 \leq l_7 + l_8~, \nonumber \\
&& |l_1 - l_4| \leq l_7 \leq l_1
 + l_4~, \ |l_2 - l_5| \leq l_8 \leq l_2 +
 l_5~, \ |l_3 - l_6| \leq l_9 \leq l_3 + l_6 ~.
\end{eqnarray}
The Wigner-$9j$ symbol obeys $72$ symmetries: 
\begin{eqnarray}
\left\{
  \begin{array}{ccc}
  l_1 & l_2 & l_3 \\
  l_4 & l_5 & l_6 \\
  l_7 & l_8 & l_9 
  \end{array}
 \right\} 
&=& (-1)^{\sum_{i = 1}^9 l_i}
\left\{
  \begin{array}{ccc}
  l_2 & l_1 & l_3 \\
  l_5 & l_4 & l_6 \\
  l_8 & l_7 & l_9 
  \end{array}
 \right\} 
= (-1)^{\sum_{i = 1}^9 l_i}
\left\{
  \begin{array}{ccc}
  l_1 & l_2 & l_3 \\
  l_7 & l_8 & l_9 \\
  l_4 & l_5 & l_6 
  \end{array}
 \right\} \nonumber \\
&&\qquad\qquad (\rm{odd \ permutation \ of \ rows \ or \ columns}) \nonumber \\
&=& \left\{
  \begin{array}{ccc}
  l_2 & l_3 & l_1 \\
  l_5 & l_6 & l_4 \\
  l_8 & l_9 & l_7 
  \end{array}
 \right\} 
= \left\{
  \begin{array}{ccc}
  l_4 & l_5 & l_6 \\
  l_7 & l_8 & l_9 \\
  l_1 & l_2 & l_3 
  \end{array}
 \right\} \nonumber \\
&&\qquad\qquad (\rm{even \ permutation \ of \ rows \ or \
 columns}) \nonumber \\
&=& \left\{
  \begin{array}{ccc}
  l_1 & l_4 & l_7 \\
  l_2 & l_5 & l_8 \\
  l_3 & l_6 & l_9 
  \end{array}
 \right\}
= \left\{
  \begin{array}{ccc}
  l_9 & l_6 & l_3 \\
  l_8 & l_5 & l_2 \\
  l_7 & l_4 & l_1 
  \end{array}
 \right\} \nonumber \\
 &&\qquad\qquad (\rm{reflection \ of \ the \ symbols})~.
\end{eqnarray}  

\section{Polarization Vector and Tensor}\label{appen:polarization}

We summarize the relations and properties of a divergenceless
polarization vector $\epsilon_a^{(\pm 1)}$ and a transverse and
traceless polarization tensor $e_{ab}^{(\pm 2)}$ \cite{2008cosm.book.....W}~.

The polarization vector with respect to a unit
vector $\hat{\mib{n}}$ is expressed using two unit vectors
$\hat{\mib{\theta}}$ and $\hat{\mib{\phi}}$ perpendicular to $\hat{\mib{n}}$ as 
\begin{eqnarray}
\epsilon_a^{(\pm 1)}(\hat{\mib{n}}) = \frac{1}{\sqrt{2}}[\hat{\theta}_a(\hat{\mib{n}})
 \pm i~\hat{\phi}_a (\hat{\mib{n}}) ]~. \label{eq:pol_vec_def}
\end{eqnarray}
This satisfies the relations: 
\begin{eqnarray}
\hat{n}^a \epsilon_a^{(\pm 1)}(\hat{\mib{n}}) &=& 0~, \nonumber \\
\epsilon^{(\pm 1) *}_a (\hat{\mib{n}}) &=& \epsilon^{(\mp 1)}_a (\hat{\mib{n}})
 = \epsilon^{(\pm 1)}_a (-\hat{\mib{n}})~, \nonumber\\
\epsilon^{(\lambda)}_a (\hat{\mib{n}}) \epsilon^{(\lambda')}_a (\hat{\mib{n}}) 
&=& \delta_{\lambda, -\lambda'} \ \ \ ({\rm for} \ \lambda, \lambda' = \pm 1)~. 
\label{eq:pol_vec_relation}
\end{eqnarray} 
By defining a rotational matrix, which transforms a unit vector parallel to the $z$-axis, namely $\hat{\mib{z}}$, to $\hat{\mib{n}}$, as
\begin{eqnarray}
S(\hat{\mib{n}}) 
\equiv \left( 
  \begin{array}{ccc}
  \cos\theta_n \cos\phi_n & -\sin\phi_n  & \sin\theta_n \cos\phi_n \\
 \cos\theta_{n} \sin\phi_{n}  &  \cos\phi_{n} & \sin\theta_{n} \sin\phi_{n} \\
 -\sin\theta_n & 0 & \cos\theta_{n}
  \end{array}
 \right)~,
\end{eqnarray}  
we specify $\hat{\mib{\theta}}$ and $\hat{\mib{\phi}}$ as
\begin{eqnarray}
\hat{\mib{\theta}}(\hat{\mib{n}}) = S(\hat{\mib{n}}) \hat{\mib{x}}~, \ \ 
 \hat{\mib{\phi}}(\hat{\mib{n}}) = S(\hat{\mib{n}}) \hat{\mib{y}}~, \label{eq:theta_phi_def}
\end{eqnarray}
where $\hat{\mib{x}}$ and $\hat{\mib{y}}$ are unit vectors parallel
to $x$- and $y$-axes.
By using Eq.~(\ref{eq:pol_vec_def}), the polarization tensor is constructed as
\begin{eqnarray}
e^{(\pm 2)}_{ab} (\hat{\mib{n}}) = \sqrt{2} \epsilon^{(\pm 1)}_a(\hat{\mib{n}})
 \epsilon^{(\pm 1)}_b(\hat{\mib{n}})~. \label{eq:pol_tens_def}
\end{eqnarray}

To utilize the polarization vector and tensor in the calculation of this paper, we need to expand Eqs.~(\ref{eq:pol_vec_def}) and (\ref{eq:pol_tens_def}) with spin spherical harmonics. 
An arbitrary unit vector is expanded with the spin-$0$ spherical harmonics as 
\begin{eqnarray}
\hat{r}_a &=& \sum_m \alpha_a^{m} Y_{1 m}(\hat{\mib{r}})~, \nonumber \\
\alpha^m_a &\equiv& \sqrt{\frac{2 \pi}{3}}
 \left(
  \begin{array}{ccc}
   -m (\delta_{m,1} + \delta_{m,-1}) \\
   i~ (\delta_{m,1} + \delta_{m,-1}) \\
   \sqrt{2} \delta_{m,0}
  \end{array}
\right)~. \label{eq:arbitrary_vec}
\end{eqnarray}
Here, note that the repeat of the index implies the summation.
The scalar product of $\alpha_a^m$ is calculated as
\begin{eqnarray}
\alpha_a^m \alpha_a^{m'} = \frac{4 \pi}{3} (-1)^m \delta_{m,-m'}~, \ \
\alpha_a^m \alpha_a^{m' *} = \frac{4 \pi}{3} \delta_{m,m'}~.
\end{eqnarray}
Through the substitution of Eq.~(\ref{eq:theta_phi_def}) into Eq.~(\ref{eq:arbitrary_vec}),
$\hat{\mib{\theta}}$ is expanded as 
\begin{eqnarray}
\hat{\theta}_a (\hat{\mib{n}}) &=& \sum_m  \alpha_a^m Y_{1m} (\hat{\mib{\theta}} (\hat{\mib{n}}))
= \sum_m  \alpha_a^m \sum_{m'} D_{m m'}^{(1) *} (S(\hat{\mib{n}})) Y_{1 m'}
 (\hat{\mib{x}}) \nonumber \\
&=& - \frac{s}{\sqrt{2}}(\delta_{s,1} + \delta_{s,-1}) \sum_m  \alpha_a^m {}_s Y_{1 m}(\hat{\mib{n}})~.
\end{eqnarray}
Here, we use the properties of the Wigner $D$-matrix as
\cite{Shiraishi:2010sm, Goldberg/etal:1967, Okamoto:2002ik, 2008cosm.book.....W}
\begin{eqnarray}
Y_{\ell m}(S(\hat{\mib{n}})\hat{\mib{x}}) &=& \sum_{m'} 
D^{(\ell) *}_{m m'} (S(\hat{\mib{n}})) Y_{\ell m'} (\hat{\mib{x}}) ~, \\
D_{ms}^{(\ell)} ( S(\hat{\mib{n}}) ) &=& 
\left[ \frac{4 \pi}{2\ell + 1} \right]^{1/2} (-1)^s
{}_{-s}Y_{\ell m}^*(\hat{\mib{n}})~.
\end{eqnarray}
In the same manner, $\hat{\mib{\phi}}$ is also calculated as
\begin{eqnarray}
\hat{\phi}_a(\hat{\mib{n}}) = \frac{i}{\sqrt{2}}(\delta_{s,1} + \delta_{s,-1})
\sum_m  \alpha_a^m {}_s Y_{1 m}(\hat{\mib{n}})~;
\end{eqnarray}
hence, the explicit form of Eq.~(\ref{eq:pol_vec_def}) is calculated as  
\begin{eqnarray}
\epsilon_a^{(\pm 1)} (\hat{\mib{n}}) 
= \mp \sum_m \alpha_a^m {}_{\pm 1} Y_{1 m} (\hat{\mib{n}})~. \label{eq:pol_vec}
\end{eqnarray}

Substituting this into Eq.~(\ref{eq:pol_tens_def}) and using the
relations of Appendix \ref{appen:wigner} and $I_{2~1~1}^{\mp 2 \pm 1 \pm
1} = \frac{3}{2 \sqrt{\pi}}$, 
the polarization tensor can also be expressed as 
\begin{eqnarray}
e^{(\pm 2)}_{ab} (\hat{\mib{n}})
&=& \frac{3}{\sqrt{2 \pi}}  
\sum_{M m_a m_b} {}_{\mp 2}Y_{2 M}^*(\hat{\mib{n}}) 
\alpha^{m_a}_{a} \alpha^{m_b}_b 
\left(
  \begin{array}{ccc}
  2 & 1 &  1\\
  M & m_a & m_b 
  \end{array}
\right)~.
\end{eqnarray}
This obeys the relations:
\begin{eqnarray}
e_{aa}^{(\pm 2)}(\hat{\mib{n}}) &=& \hat{n}_a e_{ab}^{(\pm 2)}(\hat{\mib{n}}) = 0~, \nonumber \\
e_{ab}^{(\pm 2) *}(\hat{\mib{n}}) &=& e_{ab}^{(\mp 2)}(\hat{\mib{n}}) = e_{ab}^{(\pm
2)}(- \hat{\mib{n}})~, \nonumber \\
e_{ab}^{(\lambda)}(\hat{\mib{n}}) e_{ab}^{(\lambda')}(\hat{\mib{n}}) &=& 2
\delta_{\lambda, -\lambda'} \ \ \ ({\rm for} \ \lambda, \lambda' = \pm 2)~. \label{eq:pol_tens_relation}
\end{eqnarray}

Using the projection operators as 
\begin{eqnarray}
O_a^{(0)} e^{i \mib{k}\cdot \mib{x}} &\equiv& k^{-1} \nabla_a e^{i \mib{k}\cdot \mib{x}} = i \hat{k}_a e^{i \mib{k}\cdot \mib{x}} ~, \\
O^{(0)}_{ab} e^{i \mib{k}\cdot \mib{x}} &\equiv& \left( k^{-2} \nabla_a \nabla_b + \frac{\delta_{a,b}}{3} \right) e^{i \mib{k}\cdot \mib{x}}
= \left(- \hat{k}_a \hat{k}_b + \frac{\delta_{a,b}}{3} \right) e^{i \mib{k}\cdot \mib{x}} ~, \\
O_a^{(\pm 1)} e^{i \mib{k}\cdot \mib{x}} &\equiv& - i \epsilon^{(\pm 1)}_a(\hat{\mib{k}}) e^{i \mib{k}\cdot \mib{x}} ~, \\
O^{(\pm 1)}_{ab} e^{i \mib{k}\cdot \mib{x}} 
&\equiv& k^{-1} \nabla_a O^{(\pm 1)}_b e^{i \mib{k}\cdot \mib{x}}
= \hat{k}_a \epsilon^{(\pm 1)}_b(\hat{\mib{k}}) e^{i \mib{k}\cdot \mib{x}} ~, \\
O_{ab}^{(\pm 2)} e^{i \mib{k}\cdot \mib{x}} &\equiv& e^{(\pm 2)}_{ab}(\hat{\mib{k}}) e^{i \mib{k}\cdot \mib{x}}~,
\end{eqnarray}
the arbitrary scalar, vector and tensor are decomposed into the helicity states as 
\begin{eqnarray}
\eta(\mib{k}) &=& \eta^{(0)}(\mib{k}), \label{eq:scal_decompose} \\
\omega_{a}(\mib{k}) &=& \omega^{(0)}(\mib{k}) O^{(0)}_a + \sum_{\lambda = \pm 1} \omega^{(\lambda)}(\mib{k}) O^{(\lambda)}_a~, \\
\chi_{ab}(\mib{k}) &=&  \chi^{(0)}(\mib{k}) O^{(0)}_{ab} + \sum_{\lambda = \pm 1} \chi^{(\lambda)}(\mib{k}) O^{(\lambda)}_{ab} + \sum_{\lambda = \pm 2} \chi^{(\lambda)}(\mib{k}) O^{(\lambda)}_{ab}~.
\end{eqnarray}
Then, using Eq.~(\ref{eq:pol_vec_relation}) and (\ref{eq:pol_tens_relation}), we can find the inverse formulae as
\begin{eqnarray}
\omega^{(0)}(\mib{k}) &=& -i \hat{k}_a \omega_a(\mib{k})~, \\ 
\omega^{(\pm 1)}(\mib{k}) &=& i \epsilon_{a}^{(\mp 1)}(\hat{\mib{k}}) \omega_{a}(\mib{k})~, \\
\chi^{(0)}(\mib{k}) &=& \frac{3}{2} \left(- \hat{k}_a \hat{k}_b + \frac{\delta_{a,b}}{3} \right) \chi_{ab}(\mib{k})~, \\
\chi^{(\pm 1)}(\mib{k}) &=& \hat{k}_a \epsilon_b^{(\mp 1)}(\hat{\mib{k}}) \chi_{ab}(\mib{k})~, \\
\chi^{(\pm 2)}(\mib{k}) &=& \frac{1}{2} e_{ab}^{(\mp 2)}(\hat{\mib{k}}) \chi_{ab}(\mib{k})~. \label{eq:tens_decompose} 
\end{eqnarray}


\end{document}